\documentclass[doublecol]{epl2}
\usepackage{booktabs}
\usepackage{amsmath}
\usepackage{multirow}

\title{Regional economic status inference from information flow and talent mobility}
\shorttitle{Regional economic status inference}

\author{Jun Wang\inst{1} \and Jian Gao\inst{1,2}\footnote{E-mail: gaojian08@hotmail.com} \and Jin-Hu Liu\inst{1} \and Dan Yang\inst{1} \and Tao Zhou\inst{1,2}}
\shortauthor{J. Wang \etal}

\institute{
  \inst{1} Big Data Research Center, University of Electronic Science and Technology of China, Chengdu 611731, PRC \\
  \inst{2} Institution of New Economic Development, Chengdu 610049, PRC \\
}


\abstract{Novel data has been leveraged to estimate socioeconomic status in a timely manner, however, direct comparison on the use of social relations and talent movements remains rare. In this letter, we estimate the regional economic status based on the structural features of the two networks. One is the online information flow network built on the following relations on social media, and the other is the offline talent mobility network built on the anonymized resume data of job seekers with higher education. We find that while the structural features of both networks are relevant to economic status, the talent mobility network in a relatively smaller size exhibits a stronger predictive power for the gross domestic product (GDP). In particular, a composite index of structural features can explain up to about 84\% of the variance in GDP. The result suggests future socioeconomic studies to pay more attention to the cost-effective talent mobility data.}
\begin{document}
\maketitle

\section{Introduction}
Timely estimation of social and economic status has important implications for addressing many development-related issues \cite{Schweitzer2009,Einav2014,Perc2014,Zhang2014}, such as developing policies to reduce poverty \cite{Birdsall1997}, forecasting unemployment rate \cite{Llorente2015,Yuan2016}, and optimizing strategies for economic diversification \cite{Alshamsi2018,Gao2017}. Traditional socioeconomic status inference, however, usually follows a long-time delay due to the large consumption of resources in data collection. Thanks to the technological development, novel data sources are now increasingly available for estimating socioeconomic status \cite{Blumenstock2010,Jean2016}. For example, Elvidge \etal \cite{Elvidge2009} produced a global poverty map based on the brightness of night-time lights. Gao and Zhou \cite{Gao2018} quantified regional economic complexity by analyzing firm data. Dong \etal \cite{Dong2017} measured economic activity through mining mobile phone records. Liu \etal \cite{Liu2016} inferred city-level economic status from online activities. Blumenstock \etal \cite{Blumenstock2015} predicted district-level wealth distribution based on mobile phone usages. Sobolevsky \etal \cite{Sobolevsky2017} estimated individual socioeconomic status by analyzing bank card transactions. More related works are summarized in recent reviews \cite{Einav2014,Gao2016}.

Among these works, two streams of literature are of particular interest. One stream focuses on relations between social network structure and economic status \cite{Zhang2014,Granovetter2005}. For example, Eagle \etal \cite{Eagle2010} uncovered a strong correlation between social network diversity and socioeconomic indicators, Mao \etal \cite{Mao2015} found that the ratio of in-going and out-going calls can predict a region's income level, and Holzbauer \etal \cite{Holzbauer2016} showed that cross-state long ties on social media are strongly correlated with GDP in the US. Recently, Jahani \etal \cite{Jahani2017} uncovered a strong correlation between ego-network structural diversity and individual income, and Luo \etal \cite{Luo2017} found that individuals' influence in social network is predictive to their economic status. The other stream links human mobility pattern to socioeconomic status and outcomes \cite{Carra2016,Fan2018}. Individuals with different socioeconomic status have distinct mobility patterns \cite{Yan2012,Lotero2016}, and the movement of talents is critical to economic development \cite{Kerr2016}. For example, Frias-Martinez \etal \cite{Frias2013} showed the predictive power of mobility patterns to socioeconomic status, Pappalardo \etal \cite{Pappalardo2016} found that movement diversity can well predict socioeconomic indicators, and Florez \etal \cite{Florez2018} demonstrated that a group's income increases with the diversity of commuting trips.

Most of previous works focus on either social network structures or human behavioral patterns \cite{Zhao2015}. Yet, the direct comparison between the predictive power of online social network structure and offline human mobility pattern to regional and individual socioeconomic status remain insufficient. One challenge that hinder studies towards this direction is the lack of large-scale and high-resolution online information and offline mobility data. Recently, the increasing availability of large-scale social and economic data with high spatial and temporal resolutions, such as mobile phone records \cite{Dong2017}, behavioral data \cite{Cao2018}, web-based ratings \cite{Chen2018}, public profiles \cite{Yang2018}, has made it possible to estimate socioeconomic status in a timely manner and with a relatively low cost \cite{Einav2014,Gao2017}. This provides us a chance to compare the capability of information flow and talent mobility on speculating economic status.

In this letter, we infer regional economic status from the following relations on social media and the talent movements recorded by anonymized resume data. We first build two directed and weighted networks, named online information flow network and offline talent mobility network. Then, we calculate several network structural features and link them to GDP. Results show that some features exhibit strong correlations with GDP such as the loops and outgoing spatial diversity of the information flow network and the out-strength and ingoing topological diversity of the information flow network. Overall, the talent mobility network features perform better in predicting economic status. After performing regression analysis for robustness checks, we further construct a composite index of both network structural features, which can explain up to about 84\% of the variance in GDP.

\section{Data and methods} In this section, we first introduce two large-scale online-crawled datasets for network construction, then present some measures to quantify network structures, and lastly introduce the methods applied for correlation and regression analyses.

\subsection{Data description}
\begin{figure*}
  \centering
  \includegraphics[width=0.99\textwidth]{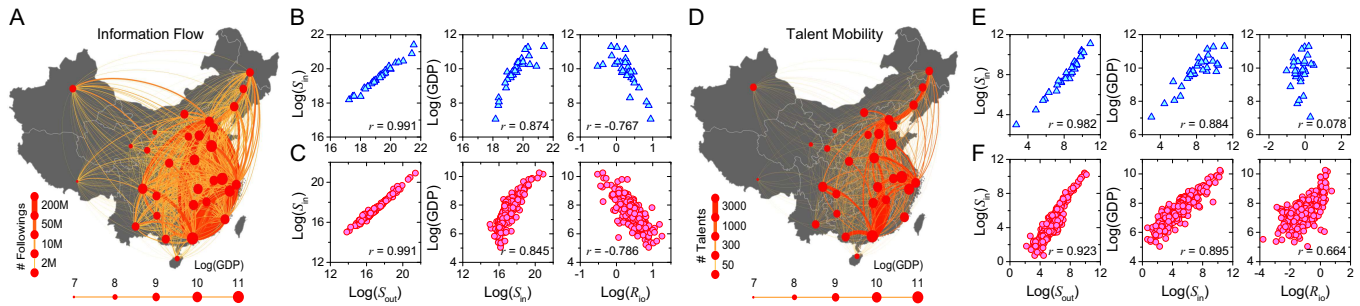}
  \caption{(Color online) Relations between network structural features and economic status. Nodes represent provinces with size showing their GDP in natural logarithmic form. (A) presents the online information flow (OIF) network with link weight being the number of followings. (B) and (C) present the relations regarding OIF at the province and city level, respectively. (D) presents the offline talent mobility (OTM) network with link weight being the number of moved talents. (E) and (F) present the relations regarding OTM at the province and city levels, respectively. The Pearson correlation coefficient $r$ is also presented.}
  \label{Fig1}
\end{figure*}

The online information flow (OIF) network is built based on the public profiles and following relations among about 433 millon users of the China's social network Weibo, which provides similar functions to Twitter. Specifically, from profiles we extract users' locations covering 336 prefecture-level cities aggregated into 31 provinces (see Ref. \cite{Liu2016} for details). Then, based on users' following relations we build the OIF network $G^{I}$ among regions (cities or provinces dependent on the resolution) and represent it by a weighted adjacent matrix $A^{I}$, whose element $a^{I}_{i,j}$ is the volume of information from region $i$ to $j$, which is roughly estimated by the number of followings from region $j$ to $i$. As users within the same region can follow each other, $G^I$ contains loops, \emph{i.e.}, $a^{I}_{i,i} \neq 0$ in $A^{I}$. Fig.~\ref{Fig1}A presents the visualization of the provincial-level OIF network and Table~\ref{Tab1} summarizes basic statistics.

\begin{table}[htb]
  \caption{Statistics of the online information flow (OIF) and the offline talent mobility (OTM) networks. The average link weight is calculated by $\langle a \rangle = \sum_{i,j}a_{i,j}/\sum_{i,j}\delta_{i,j}$, where $\delta_{i,j}=1$ if there is a link from node $i$ to $j$, and $\delta_{i,j}=0$ if otherwise.}
    \scriptsize
    \begin{tabular*}{0.48\textwidth}{llcccc}
    \toprule
    Network & Resolution & \# Regions & \# Links  & $\langle a \rangle$   \\
    \midrule
    \multirow{2}[1]{*}{OIF} & Province & 31    & 961     & $1.277 \times 10^{7}$ \\
     & City  & 336   & 112,896      & $1.087\times 10^{5}$  \\
    \multirow{2}[1]{*}{OTM} & Province & 31    & 818  & 347.7  \\
     & City  & 287   & 9,746  & 29.18 \\
    \bottomrule
    \end{tabular*}
  \label{Tab1}
\end{table}

The offline talent mobility (OTM) network is built based on the self-reported resume data of about 142 thousand anonymized Chinese job seekers with higher education (see Ref. \cite{Yang2018} for details). Specifically, we roughly estimate the flow of talents among regions based on the movements of job seekers from birth city to living city in career development and from living city to expected city in job hunting. The resume data covers 287 prefecture-level cities aggregated into 31 provinces. Notice that, some cities are isolated due to sparsity, and only cities remaining in the giant connected network are counted. The directed and weighted OTM network $G^{T}$ can also be represented by a weighted adjacent matrix $A^{T}$, whose element $a^{T}_{i,j}$ is the number of talents moved from region $i$ to $j$. Similarly, $G^{T}$ contains loops. Fig.~\ref{Fig1}D visualizes the provincial-level OTM network and Table~\ref{Tab1} summarizes basic statistics.

Some macro economic data at the province and city levels are collected respectively from the official books entitled ``China Statistical Yearbook (2017)'' and ``China City Statistical Yearbook (2017)'' released by the National Bureau of Statistics of China. Due to the time-consuming statistics, these books provide data with one year-lag, namely, for the year 2016. We have successfully collected GDP of 31 provinces and 290 prefecture-level cities while failed for the rest 46 cities due to the missing data. The unit of GDP data is 10,000 RMB (about 1,500 USD).

\subsection{Structural features}
Considering a network $G$ with a weighted adjacency matrix $A$, we first calculate three direct structural features, namely, $S_{out}$, $S_{in}$, and $S_{loop}$ \cite{Barabasi2016}. Specifically, for a region $i$, $S_{out}(i)=\sum_{j}a_{i,j}$ sums the weights of outgoing links, $S_{in}(i)=\sum_{j}a_{j,i}$ sums the weights of ingoing links, and $S_{loop}(i)=a_{i,i}$ is the weight of the self-loop link. Then, we calculate three relative structural features, namely, $R_{io}$, $R_{lo}$ and $R_{li}$. Specifically, $R_{io}(i)= S_{in}(i)/S_{out}(i)$ measures the rates of local information/talent retention. $R_{lo}(i)=S_{loop}(i)/S_{out}(i)$ measures information/talent drain, where $R_{lo}=0$ and $R_{lo}=1$ mean all information/talents are drained and kept, respectively. $R_{li}(i)=S_{loop}(i)/S_{in}(i)$ measures information/talent gain, where $R_{li}=0$ means new information/talents are gained and $R_{li}=1$ means previous information/talents are kept.

Moreover, we quantify diversity by calculating four network structural features: two topological diversity measures ($H_{out}$ and $H_{in}$) and two spatial diversity measures ($D_{out}$ and $D_{in}$) \cite{Eagle2010}. Specifically, the ingoing and outgoing topological diversity of a region is defined by the Shannon entropy associated with the information/talent flow into and out of the region, respectively. Formally, the outgoing topological diversity $H_{out}(i)$ for region $i$ is given by
\begin{equation}
H_{out}(i) = -\sum_{j\neq i}p_{i,j}\log(p_{i,j}),
\end{equation}
where $p_{i,j}=a_{i,j}/\sum_{j}a_{i,j}$. The ingoing spatial diversity $D_{out}(i)$ for region $i$ is calculated by normalizing $H_{out}(i)$ using the number of involved regions. Mathematically,
\begin{equation}
D_{out}(i) = \frac{H_{out}(i)}{\log(k_{out}(i))},
\end{equation}
where $k_{out}(i)$ is the out-degree of region $i$. Analogously, the ingoing topological diversity $H_{in}(i)$ for region $i$ is defined in the similar manner, by
\begin{equation}
H_{in}(i) = -\sum_{j\neq i}p_{j,i}\log(p_{j,i}),
\end{equation}
where $p_{j,i}=a_{j,i}/\sum_{j}a_{j,i}$. The ingoing spatial diversity $D_{in}(i)$ for region $i$ is calculated by normalizing $H_{in}(i)$ using the number of involved regions, as
\begin{equation}
D_{in}(i) = \frac{H_{in}(i)}{\log(k_{in}(i))},
\end{equation}
where $k_{in}(i)$ is the in-degree of region $i$.

\subsection{Analytical methods}
To exploit the relations between structural features and GDP, we perform both correlation analysis and regression analysis. The Pearson correlation coefficient $r$ is used to quantify the linear correlation between two variables. The value $r$ is in the range $[-1,1]$, from negative to positive correlation. The ordinary least squares (OLS) model is employed to regress GDP against structural features. The estimated equation is given by
\begin{equation}
\begin{aligned}
\log(GDP) & = \beta_{0} + \beta_{1}S_{out} + \beta_{2}S_{in} + \beta_{3}R_{io}  \\
&  + \beta_{4}S_{loop} + \beta_{5}R_{lo} + \beta_{6}R_{li} + \beta_{7}H_{out}  \\
&  + \beta_{8}H_{in} + \beta_{9}D_{out} + \beta_{10}D_{in} + \varepsilon ,
\end{aligned}
\end{equation}
where the structural variables are in the logarithmic form expect for the diversity measures, $\{ \beta_0, \beta_1, \cdots, \beta_{10} \}$ are regression coefficients of variables, and $\varepsilon$ is the error term.

\section{Results}
\label{Sec3}
In this section, we first analyze correlations between simple structural features and GDP, then summarize correlations between diversity-related features and GDP, and finally perform some robustness checks using regression models, based on which a composite index is further constructed to explore the prediction accuracy.

\subsection{Correlation between simple features and GDP}

The visualizations of province-level online information flow (OIF) and offline talent mobility (OTM) networks are presented in Fig.~\ref{Fig1}A and \ref{Fig1}D, in which the direct link weights are the numbers of followings and talents from origin to target provinces, respectively. For OIF, Fig.~\ref{Fig1}B and \ref{Fig1}C (Left) present the relations between $S_{out}$ and $S_{in}$ at the province and city levels, respectively. We find that $S_{out}$ and $S_{in}$ are perfectly correlated with each other, as suggested by $r \approx 0.99$ at both resolutions. In contrast, as shown in Fig.~\ref{Fig1}E and \ref{Fig1}F (Left), the correlations between $S_{out}$ and $S_{in}$ for OTM are relatively weaker, suggesting the unbalance of talent flows into and out of regions.

The volume of information and talent flows can be relevant to a region's economic status. For OIF, Fig.~\ref{Fig1}B and \ref{Fig1}C (Middle) present the relations between $S_{in}$ and GDP at the province and city levels, respectively. We notice that $S_{in}$ exhibits a high correlation ($r \approx 0.86$) with GDP. Fig.~\ref{Fig1}E and \ref{Fig1}F (Middle) present the similar trend for OTM, while the correlations ($r \approx 0.89$) are stronger at both resolutions. The ratio of ingoing and outgoing flows can also be linked to economic status. For OIF, Fig.~\ref{Fig1}B and \ref{Fig1}C (Right) present the relations between $R_{io}$ and GDP, where we find negative correlations ($r \approx -0.78$) at both resolutions. This suggests that developed regions spread information better. As presented by Fig.~\ref{Fig1}E and \ref{Fig1}F (Right) for OTM, however, we find a positive correlation ($r\approx 0.66$) only at the province level.

These results suggest that attractiveness for talents in fine-grained regions reflects economic status better. This observation may be originated from the inequality of regional economic development. For instance, China faces seriously unbalanced regional economic development, where more developed cities usually have talent gain, while less developed cities may have talent drain. This unbalanced talent mobility and economic development at the city level may result in the positive correlations. However, such correlation can be diminished at the aggregated province level as a province can have multiple cities with different social and economic status, and talents can move among cities located in the same province.

\begin{figure}
  \centering
  \includegraphics[width=0.48\textwidth]{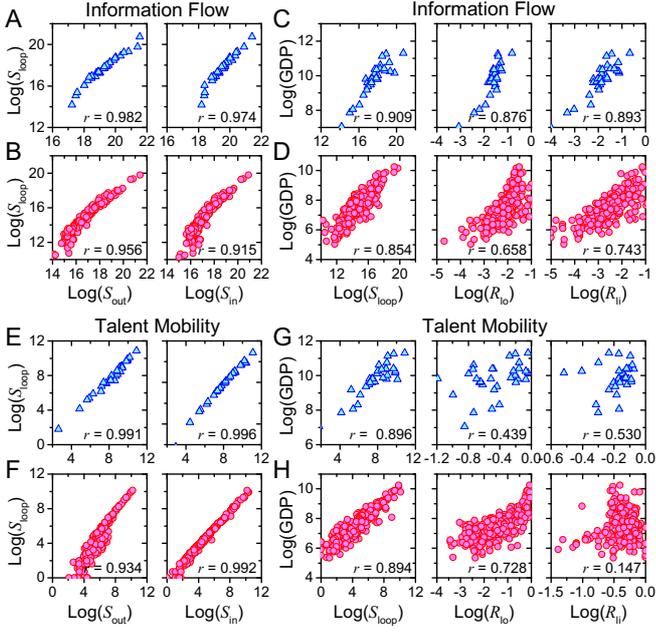}
  \caption{Relations between network loops and economic status. (A) and (B) present relations between loops and strengths of the online information flow (OIF) network at the province and city levels, respectively. Analogously, (E) and (F) present relations for the offline talent mobility (OTM) network. (C) and (D) present relations between loop-related measures of OIF and GDP at the province and city levels, respectively. Analogously, (G) and (H) present relations for OTM. The Pearson correlation coefficient $r$ is also presented.}
  \label{Fig2}
\end{figure}

The strength of loops ($S_{loops}$) in the OIF and OTM networks suggest the retention of local information and talents, respectively. For OIF, Fig.~\ref{Fig2}A and \ref{Fig2}B present how $S_{loops}$ is related to $S_{out}$ (Left) and $S_{in}$ (Right) at province and city levels, respectively. Similarly, Fig.~\ref{Fig2}E and \ref{Fig2}F present the relations for OTM. Overall, we find that loops are perfectly correlated ($r\approx 0.99$) with strengths. Further, we explore how information and talent retentions are linked to economic status by calculating correlations between GDP and three loop-related features, namely, $S_{loop}$, $R_{lo}$, and $R_{li}$. For OIF, we find from Fig.~\ref{Fig2}C and \ref{Fig2}D that GDP is positively correlated with all the three features, and $S_{loop}$ exhibits the strongest correlation ($r\approx 0.91$) at the province level. Similar results hold for OTM as shown in Fig.~\ref{Fig2}G and \ref{Fig2}H, and $S_{loops}$ has a high correlation ($r\approx 0.90$) with GDP at both resolutions. These results suggest the predictive power of local information and talent retentions for regional economic status.

\subsection{Correlation between diversity features and GDP}

\begin{figure}
  \centering
  \includegraphics[width=0.47\textwidth]{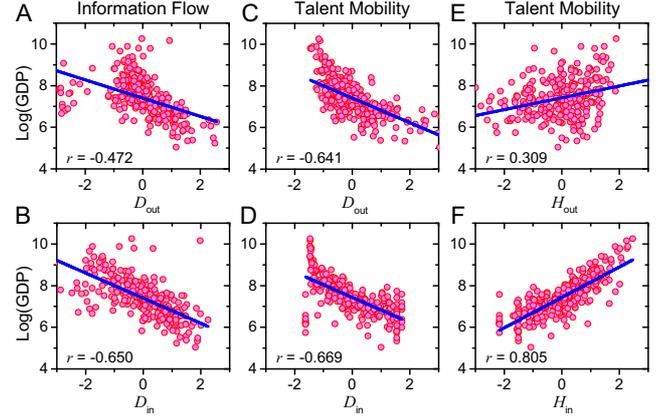}
  \caption{Relations between diversity-related features and economic status. (A) and (B) present how GDP is related to $D_{out}$ and $D_{in}$ of the online information flow (OIF) network, respectively. Analogously, (C) and (D) present relations for the offline talent mobility (OTM) network. (E) and (F) present how GDP is related to $H_{out}$ and $H_{in}$ of OTM, respectively. Lines present the linear fits, and $r$ is the Pearson correlation coefficient.}
  \label{Fig3}
\end{figure}

We explore relations between GDP and two diversity-related features, namely, spatial diversity ($D$) and topological diversity ($H$). For OIF, Fig.~\ref{Fig3}A and \ref{Fig3}B present how GDP is related to the outgoing ($D_{out}$) and ingoing ($D_{in}$) spatial diversities, respectively. We observe strong negative correlations for both cases, specifically, $r=-0.472$ for $D_{out}$ and $r=-0.650$ for $D_{in}$. As shown in Fig.~\ref{Fig3}C and \ref{Fig3}D, while similar observations hold for OTM, the correlations ($r\approx -0.66$) are stronger. In particular, we notice that $D_{in}$ has a stronger correlation with GDP for both networks. Previous study based on the UK communications showed that social network spatial diversity is positively correlated with community-level development \cite{Eagle2010}, however, our results based on both the OIT and OTM networks in China suggest spatial diversities as negative predictors of regional economic status.

The topological diversity is equal to the spatial diversity for OIF as it is fully connected. Thereby, only for OTM we present how GDP is related to topological diversities $H_{out}$ and $H_{in}$ in Fig.~\ref{Fig3}E and \ref{Fig3}F, respectively. We find that the correlation $r=0.805$ between $H_{in}$ and GDP is significantly larger than the correlation $r=0.309$ between $H_{out}$ and GDP, showing that $H_{in}$ is a more relevant feature to economic status. In summary, we find that $D$ and $H$ of OIF and $D$ of OTM are negative predictors of GDP, while $H$ of OTM is positively correlated with GDP.

\subsection{Regression analysis and composite index}

\begin{figure}
  \centering
  \includegraphics[width=0.47\textwidth]{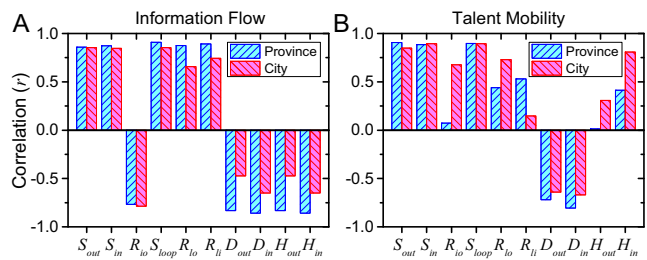}
  \caption{(Color online) Pearson correlations between structural features and GDP. (A) and (B) present correlations between GDP and structural features of the online information flow network and the offline talent mobility network, respectively.}
  \label{Fig4}
\end{figure}

The Pearson correlations between structural features and economic development (GDP) are summarized in Fig.~\ref{Fig4}. As shown in Fig.~\ref{Fig4}A for OIF, simple structural features except $R_{io}$ have strongly positive correlations with GDP, while diversity-related features exhibit strongly negative correlations. Moreover, network structural features are more relevant to economic status at the province level than at the city level. In particular, the most relevant features are loops and diversities at the province level as well as strengths and loops at the city level. As presented in Fig.~\ref{Fig4}B for OTM, the most relevant features are $S_{out}$, $S_{in}$ and $S_{loop}$ at both the province and city levels.

We further perform some robustness checks by employing the ordinary least squares (OLS) model to regress GDP against structural features at the city level. Table~\ref{Tab2} summaries the regression results. As shown in columns (1) and (2), the model including the OIF and the OTM network structural features can explain up to 76.2\% and 80.2\% of the variance in GDP, respectively. In particular, we notice that $D_{out}$ and $D_{in}$ of OIF are respectively significantly positive and negative predictors of GDP, while only $H_{out}$ of OTM is a positive predictor of GDP. Including both network structural features in column (3), where column (3-1) and column (3-2) correspond respectively to OIF and OTM, the variance in GDP can be explained by up to 83.2\%. We additionally find that loops are the features that can best explain the variance in GDP with the adjusted $R^2 = 74.3\%$ for OIF and $R^2 = 79.8\%$ for OTM. These results confirm that the OTM network structural features are more predictive to regional economic development.

Based on the regression analysis, we construct a composite index of network structural features for the best prediction of regional economic status. Specifically, the composite index is calculated through weighting structural features by their regression coefficients. Formally, the composite index $CI(i)$ for region $i$ is given by
\begin{equation}
CI(i) = \sum_{j=1}^{10} \beta^{I}_{j} M^{I}_{j,i} + \sum_{j=1}^{10} \beta^{T}_{j} M^{T}_{j,i}.
\end{equation}
$M = \{ \vec{S}_{out}, \vec{S}_{in}, \vec{R}_{io}, \vec{S}_{loop}, \vec{R}_{lo}, \vec{R}_{li}, \vec{D}_{out}, \vec{D}_{in}, \vec{H}_{out}, \vec{H}_{in} \}$ is the ten vectors of network structural features, and $\vec{\beta} = \{ \beta_{1}, \cdots, \beta_{10} \}$ is the vector of corresponding regression coefficients as shown in Table~\ref{Tab2}. Specifically, $M^I$ and $\beta^I$ are for OIF, and $M^T$ and $\beta^T$ are for OTM. All network structural features are standardized by the $z$-score \cite{Kreyszig1979} before constructing the composite index.

\begin{figure}[!t]
  \centering
  \includegraphics[width=0.47\textwidth]{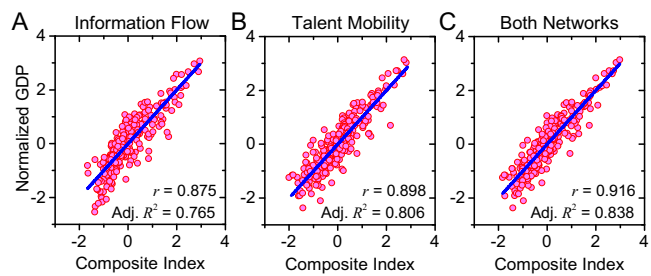}
  \caption{Predictive power of the composite index for economic development. (A) and (B) present how GDP is related to the composite index of online information flow and offline talent mobility network structural features. (C) presents the relation between GDP and the composite index of both networks. The Pearson correlation coefficient $r$ is presented. Line is the linear fit, and the adjusted $R^2$ shows its explanatory power.}
  \label{Fig5}
\end{figure}

The correlations between the composite index and normalized GDP at the city level are presented in Fig.~\ref{Fig5}A and \ref{Fig5}B for OIF and OTM, respectively. For both networks, we find that GDP is strongly and positively correlated with the composite index. In particular, the composite index of OTM exhibits a slightly larger correlation ($r=0.898$) with GDP than the one of OIF ($r=0.875$). The composite index of OIF and OTM can explain 76.5\% and 80.6\% of the variance in GDP, respectively. These observations suggest strong predictive powers of information and talent flows for regional economic development. Further, we construct a composite index using the structural features of both networks. As shown in Fig.~\ref{Fig5}C, the composite index has the largest correlation ($r=0.916$) with GDP, and it can explain up to 83.8\% the variance in GDP. The result shows that combining network features of information flow and talent mobility can enhance the performance of economic status inference.

\begin{table}[!t]
  \centering
  \footnotesize
  \caption{The predictive power of network structural features for economic development. The OLS model is used to regress $\log(GDP)$ against network structural features. All variables are standardized by the $z$-score before being included. The symbol ``$--$'' marks omitted variables due to high collinearity that are automatically identified by the regression model. Significant level: $^{*}p<0.1$, $^{**}p<0.05$, and $^{***}p<0.01$.}
    \begin{tabular*}{0.48\textwidth}{lcccc}
    \toprule
    \multirow{2}[3]{*}{Variables} & \multicolumn{4}{c}{OLS Model} \\
    \cmidrule{2-5}
          &(1)   & (2)  & (3-1) & (3-2) \\
    \midrule
    $S_{out}$  & 0.823$^{***}$& 0.587$^{***}$ & $--$     & 0.266$^{***}$ \\
    $S_{in}$ & $--$  &   $--$     & 0.363$^{***}$ &  $--$  \\
    $R_{io}$ & 0.217$^{**}$ & 0.300$^{***}$ & 0.051 & 0.128$^{**}$ \\
    $S_{loop}$ &   $--$     &   $--$  &  $--$      &  $--$  \\
    $R_{lo}$ & 0.216$^{***}$ &  $--$  & 0.087$^{*}$ &  $--$  \\
    $R_{li}$ &  $--$      & 0.041 &  $--$      & 0.059$^{**}$ \\
    $D_{out}$ & 0.208$^{***}$ & $-$0.010 & 0.203$^{***}$ & $-$0.096 \\
    $D_{in}$ & $-$0.291$^{***}$ & 0.013 & $-$0.192$^{***}$ & 0.010 \\
    $H_{out}$ &  $--$      & 0.067$^{**}$ &  $--$      & 0.016 \\
    $H_{in}$ &  $--$      & 0.103 &  $--$      & 0.120$^{*}$ \\
    \midrule
    Obs.  & 290   & 280   & \multicolumn{2}{c}{280} \\
    Adj. $R^2$ & 0.762 & 0.802 & \multicolumn{2}{c}{0.832} \\
    \bottomrule
    \end{tabular*}
  \label{Tab2}
\end{table}

\section{Conclusion and discussions}
\label{Sec4}
In summary, we have explored the inference of regional economic status from the online information flow network and the offline talent mobility network. The former was built on the following relations among about 433 million social media users, and the latter was built on the self-reported resume data of over 142 thousand job seekers with higher education. After performing the correlation analysis, we found that strengths of both networks have strongly positive correlations with GDP, and the loop-related network features are the most relevant. Moreover, we uncovered the negative correlations between GDP and the spatial diversities for both networks, while the topological diversities of the talent mobility network are positively correlated with GDP. Interestingly, we found that the talent mobility network features exhibit a stronger predictive power for GDP although it covers only about 1/3000 people in comparison with the information flow network. This suggests a more cost-effective way to infer economic status by leveraging some relative small-scale offline talent mobility data

The correlations between GDP and the information flow network structural features diminish at the fine-grained resolution. In particular, we observed negative correlations between spatial diversities and GDP, which is different from the previous finding \cite{Eagle2010}. Whether this inconsistency is originated from the inequality and complexity of China's regional development \cite{Gao2018} remains an open issue. Through the regression analysis, we found that the significant predictors of GDP are out-strength, ratios of loops and spatial diversities of the information flow network as well as the out-strength, loops and outgoing topological diversities of the talent mobility network. Based on the regression results, we further constructed a composite index of both network structural features that can explain up to about 84\% of the variance in GDP. The result suggests a way of improving economic status inference through combining different network information.

The presented results should be interpreted in the light of some limitations on the data and analytical methods, which ask for further explorations. The estimation of information flow was solely based on social media, where taking into account other information exchange channels such as online chats \cite{Guille2013} and mobile communications \cite{Onnela2007,Wang2016} would help. The resume data covers a relatively small sample, where adding other large-scale data from human resource services \cite{Yang2018}, academic publishers \cite{Deville2014} and formal talent markets \cite{Murphy1991} will be an improvement. Recent available large-scale and high spatio-temporal data would advance studies on comparing the predictive power of different data sources on inferring socioeconomic status. Moreover, a limited number of structural features were considered, where many network ranking indicators \cite{Lu2016} can also be considered. In addition, it would be interesting to apply some variant models to predict and validate regional and temporal change of GDP based on time-windowed past GDP and network data, and we leave this for future work when data are available. Keeping these aforementioned limitations in mind, we hope our work will spark further studies on economic status inference from the aspects of both information flow and talent mobility.

\acknowledgments
The authors acknowledge Hao Chen, Jing-Yi Liao, Zhong-Zheng Peng, Zhi-Hai Rong and Jun-Ming Shao for helpful discussions and Rui-Tong Wang for processing the raw data files. This work was partially supported by the National Natural Science Foundation of China (Grant Nos. 61433014, 61603074, 61673086, and 61703074).

\end{document}